\documentclass[12pt]{book}

\usepackage[dvips]{graphicx,color}
\usepackage{makeidx,physics,cosmology}

\makeauthorindex \makeindex

\BookTitle{Frontier in Astroparticle Physics and Cosmology}
\CopyRight{\copyright 2004 by Universal Academy Press, Inc.}

\begin{document}

\BookTitle{\itshape Frontier in Astroparticle Physics and
Cosmology} \CopyRight{\copyright 2004 by Universal Academy Press,
Inc.}
\pagenumbering{arabic}

\chapter{Rotating Black Holes at Future Colliders} 

\author{%
Daisuke Ida\\ 
{\it Department of Physics, Tokyo Institute of Technology, Tokyo
152-8551, Japan}\\
Kin-ya Oda \\
{\it Physik Dept.\ T30e, TU M\"unchen,  James Franck Str., D-85748
Garching, Germany}\\
Seong Chan Park\\ 
{\it Korea Institute for Advanced Study (KIAS), Seoul 130-012,
Korea}}

\AuthorContents{D.\ Ida, K.-y.\ Oda and S .\ C.\ Park}

\AuthorIndex{Ida}{D.} 
\AuthorIndex{Oda}{K.-y.}
\AuthorIndex{Park}{S.C.}

\section*{Abstract}

 We consider the production and decay of TeV-scale black holes. Evaluation of
the production cross section of higher dimensional rotating black
holes is made. The master field equation for general spin-$s$
fields confined on brane world is derived. For five-dimensional
(Randall-Sundrum) black holes, we obtain analytic formulae for the
greybody factors in low frequency expansion.

\section{Introduction}
The scattering process of two particles at CM energies in the
trans-Planck domain, is well calculable using known laws of
physics, because gravitational interaction dominates over all
other interactions. Non-trivial quantum gravitational (or string/M
theoretical) phenomena are well behind the horizon. If the impact
parameter is less than the black hole radius corresponding to the
CM energy then one naturally expects a black hole to form.
When nature realizes TeV scale gravity
scenario~\cite{Arkani-Hamed:1998rs}, one of the most intriguing
prediction would be copious production of TeV sized black holes at
near future particle colliders
\cite{Banks:1999gd,Giddings:2001bu}.
The production cross section of black hole in the higher
dimensional case was obtained in ref.~\cite{Ida:2002ez} by taking
angular momenta into account  and the result has been numerically
proved in refs.~\cite{Eardley:2002re}(See also Ref.
\cite{Park:2001xc} and \cite{Anchordoqui:2001cg}.).
Once produced, black holes lose its masses and angular momenta
through the Hawking radiation \cite{Hawking:1975sw}. The Hawking
radiation is determined for each mode by the greybody factor,
i.e.\ the absorption probability  of an in\-com\-ing wave of the
corresponding mode.
The master equation for general brane-fields with arbitrary
spin-$s$ was obtained in ref.~\cite{Ida:2002ez} for rotating black
holes in higher dimensional spacetime and its non-rotating limit
was confirmed in ref.~\cite{Harris:2003eg}.
Analytic expressions of greybody factors for rotating black holes
were obtained in five dimensional (Randall-Sundrum) case in
ref.~\cite{Ida:2002ez} and also for non-rotating limit in the
series of papers ~\cite{Kanti:2002ge}.

\section{Production of rotation black holes}

Let us imagine a collision of two massless particles with finite
impact parameter $b$ and the center of mass (CM) energy
$\sqrt{s}=M_i$ so that each particle has energy $M_i/2$ in the CM
frame.
The initial angular momentum before collision is $J_i=bM_i/2$ in
the CM frame. Suppose that a black hole forms whenever the initial
two particles can be wrapped inside the event horizon of the black
hole with the mass $M=M_i$ and angular momentum $J=J_i$ , i.e.,
when
\begin{eqnarray}
b &<& 2r_h(M,J)=2r_h(M_i,bM_i/2), \label{eq:our_condition}
\end{eqnarray}
where $r_h(M,J)$ is the size of event-horizon for given energy $M$
and angular momentum $J$\cite{Myers:un}. Since the right hand side
is monotonically decreasing function of $b$, there is maximum
value $b_{max}$ which saturates the
inequality~(\ref{eq:our_condition}). The production cross section
is now given by
\begin{eqnarray}
\sigma(M) &=&\pi b_{max}^2 \nonumber \\
   &=&4\left[1+\left(\frac{n+2}{2}\right)^2\right]^{-2/(n+1)}\,\pi r_S(M)^2
   \nonumber\\
&=&F\,\pi r_S(M)^2.
\end{eqnarray}
The form factor $F$ is summarized as
\begin{eqnarray}
\begin{array}{c|cccccccc}
n  & 1 & 2 & 3 & 4 & 5 & 6 & 7 \\
\hline F_\mathrm{NY}~
   & 1.084 & 1.341 & 1.515 & 1.642 & 1.741 & 1.819 & 1.883\\
F_\mathrm{Our}
   & 1.231 & 1.368 & 1.486 & 1.592 & 1.690 & 1.780 & 1.863
\end{array}. \label{eq:Rsq_table} \nonumber
\end{eqnarray}
and the result fits the numerical result of $\sigma(M)$ with full
consideration of the general relativity by Yoshino and
Nambu~\cite{Yoshino:2002tx} within the accuracy less than 1.5\%
for $n\geq 2$ and 6.5\% for $n=1$. This result implies that  we
would underestimate the production cross section of black holes if
we did not take the angular momentum into account and that it
becomes more significant for higher dimensions.

\section{Decay of rotating black holes}
Black hole decays through the Hawking radiation. The Hawking
radiation is not exactly thermal but modified by the so-called
greybody factor which could be obtained by solving wave equations
under the black hole background metric. For the higher dimensional
rotating black holes, the brane field equations for massless spin
$s$ field was derived utilizing the Newman-Penrose
formalism~\cite{Newman:1961qr}. By the standard decomposition
$\Phi_s=R_s(r)S(\vartheta)e^{-i\omega t+im\varphi}$, the master
wave equation is given as:
\begin{eqnarray}
{1\over\sin\vartheta}{d\over d\vartheta}\left(\sin\vartheta {d
S\over d\vartheta}\right) +[ (s-a\omega\cos\vartheta)^2
-(s\cot\vartheta+m\csc\vartheta)^2 +\hat{A}]S=0,
\label{eq:angular}\\
\Delta^{-s}{d\over dr}\left(\Delta^{s+1}{dR\over dr}\right)
+\Biggl[ {K^2\over\Delta} +s\left( 4i\omega r -i{
\Delta_{,rr}K\over\Delta} +\Delta,_{rr}-2\right)
\nonumber\\
+2ma\omega-a^2\omega^2-A \Biggl]R=0, \label{eq:Teukolsky}
\end{eqnarray}
where $ K=(r^2+a^2)\omega-ma$ and $\hat{A}=-s(s-1)+A$. The
solution of the angular equation is known as spin-weighted
spheroidal harmonics ${}_sS_{lm}$. We solved the radial equation
for $n=1$ Randall-Sundrum black hole with the low frequency
expansion. Here we outline our procedure: First we obtain the
``near horizon'' and ``far field'' solutions in the corresponding
limits; Then we match these two solutions at the ``overlapping
region'' in which both limits are consistently satisfied; Finally
we impose the ``purely ingoing'' boundary condition at the near
horizon side and then read the coefficients of ``outgoing'' and
``ingoing'' modes at the far field side. The ratio of these two
coefficients can be translated into the absorption probability of
the mode, which is nothing but the greybody factor itself.
Finally, the greybody factor $\Gamma$ (=the absorption
probability) could be written as follows.
\begin{eqnarray}
\Gamma =1-\left|\frac{Y_{\rm out}
  Z_{\rm out}}{Y_{\rm in}Z_{\rm in}}\right|
=1-\left|\frac{1-C}{1+C}\right|^2, \label{eq:Page_trick}
\end{eqnarray}
where
\begin{eqnarray}
C=\frac{(4i\tilde{\omega})^{2l+1}}{4}\left(\frac{(l+s)!(l-s)!}{(2l)!(2l+1)!}\right)^2
   \left(-iQ-l\right)_{2l+1},
\end{eqnarray}
with $(\alpha)_n=\prod_{n'=1}^n(\alpha+n'-1)$ being the
Pochhammer's symbol and some useful dimensionless quantities $\xi
= \frac{r-r_h}{r_h}, \hspace{0.3cm}   \tilde{\omega}= r_h \omega$
and $ Q    = \frac{\omega-m \Omega}{2\pi T}$.  We note that the
so-called s-wave dominance is maximally violated for fermion and
vector fields since there are no $l=0$ modes for them.

\section{Summary}
We have studied theoretical aspects of the rotating black hole
production and evaporation. For production, we present an
estimation of the geometrical cross
section 
up to unknown mass and angular momentum loss in the balding phase.
Our result of the maximum impact parameter $b_{max}$ is in good
agreement with the numerical result by 
Yoshino and Nambu when the number of extra dimensions is $n\geq 1$
(i.e.\ within 6.5\% when $n=1$ and 1.5\% when $n\geq 2$). Relying
on this agreement, we obtain the (differential) cross section for
a given mass and an angular momentum. The result shows that black
holes tend to be produced with large angular momenta.  For
evaporation, we first derive the master equation for brane fields
for general spin and for an arbitrary number of extra dimensions.
We show that the equations are separable into radial and angular
parts. From these equations, we obtain the greybody factors for
brane fields with general spin for the five-dimensions ($n=1$)
Kerr black hole within the low-frequency expansion.


\end{document}